# Optimal Power Flow in Renewable-Integrated Power Systems: A Comprehensive Review

Zigang Chen [1]

[1] School of Electrical and Information Engineering, Beihua University, Jilin 132013, China

**Abstract**：This paper explores the integration of renewable energy sources into power systems, highlighting the resulting complexities such as variability and intermittency that challenge traditional power flow dynamics. We delve into innovative Optimal Power Flow (OPF) strategies designed to manage the unpredictability of renewable sources while ensuring economically viable and stable grid operations. A thorough review of state-of-the-art OPF algorithms, particularly those that enhance systems with substantial renewable integration, is presented. The discussion spans fundamental OPF principles, adaptations to renewable energies, and categorization of the latest advancements in areas such as energy uncertainty management, energy storage integration, linearization techniques application, and data-driven strategy utilization. Each sector's application benefits and limitations are critically analyzed. The paper concludes by pinpointing ongoing challenges and suggesting future research trajectories to foster adaptable and robust power system operations in the renewable-dominant energy era.

## 1. Introduction

Under the 'Dual Carbon' strategy, the nation has accelerated the construction of a new type of power system predominantly powered by renewable energy. As of the end of April 2024, the total installed generation capacity nationwide has exceeded 3 billion kilowatts, with solar power installations accounting for about 670 million kilowatts—an increase of 52.4% year-over-year—and wind power installations at about 460 million kilowatts, marking a 20.6% increase from the previous year [1]. The rising penetration rate of renewable energy has transformed the distribution network's flow into a bidirectional, complex mesh topology. The intricate network structure and flow characteristics have complicated the distribution network's flow and optimal power flow issues. Additionally, the intermittent and fluctuating nature of renewable power generation increases the overload rate of the grid, imposing higher demands for economically stable grid operation [2-3]. Addressing the grid optimization flow issues considering the integration of new energy sources is crucial for grid optimization scheduling.

Optimal Power Flow (OPF) distribution is a highly uncertain nonlinear optimization problem that requires adjusting various control measures within the grid to ensure safe and secure operations while achieving predefined optimization objectives [4-5]. OPF plays a significant role in maintaining safe operations, economic scheduling, and reliability in complex power systems. Traditional OPF algorithms include the simplified gradient method [6], Newton's method[7], the interior-point method[8], and decoupling methods. However, these classical methods require precise mathematical modeling, and building complex models might not always provide the accuracy needed for practical control. In recent years, heuristic algorithms, particularly genetic algorithms and simulated annealing, have been extensively developed and have shown effective control in large-scale power system OPF calculations.

Given this backdrop, this paper explores renewable energy OPF modeling methods, delves into the impact of renewable energy on power system OPF calculations, and discusses the improved strategies for OPF under the volatility and uncertainty of renewable energy. It summarizes the advantages and disadvantages of various improvement methods and identifies the unresolved issues and potential research directions in OPF analysis and research moving forward.

## 2. OPF Model

The mathematical model for OPF is as follows [9-11]:

$$\begin{cases} \min f = f(x) \\ g(x) = 0 \\ h(x) \leqslant 0 \end{cases} \quad (1)$$

where $f$ denotes an objective function, which is typically the coal consumption cost of the generating units, but can also represent system network losses, economic benefits of reactive power compensation, etc.; $g(x)$ stands for the power flow equation, and $h(x)$ represents the inequity constraint [12-15].

The OPF optimization that includes wind farms must satisfy the basic power flow equations, with equality constraints as follows [16-17]:

$$\Delta P_i = P_{Gi} - P_{Li} = \sum_{j=1}^{N} V_i V_j (G_{ij} \cos \delta_{ij} + B_{ij} \sin \delta_{ij}) \quad (2)$$

$$\Delta Q_i = Q_{Gi} - Q_{Li} = \sum_{j=1}^{N} V_i V_j (G_{ij} \sin \delta_{ij} - B_{ij} \cos \delta_{ij}) \quad (3)$$

where $N$ is the number of nodes in the system; $V_i$ and $V_j$ are the voltage magnitudes at nodes $i$ and $j$, respectively; $P_{Gi}$ and $Q_{Gi}$ are the active and reactive power outputs of the generator at node $i$; $P_{Li}$ and $Q_{Li}$ are the active and reactive power loads at node $i$; $\delta_{ij}$ is the phase angle difference between nodes $i$ and $j$; $G_{ij}$ and $B_{ij}$ are the real and imaginary parts of the mutual admittance between nodes $i$ and $j$, respectively.

The active power inequality constraints for the generators are [18-19]:

$$P_{Gi}^{\min} \leqslant P_{Gi} \leqslant P_{Gi}^{\max} \quad (4)$$

The reactive power inequality constraints for the generators are [20]:

$$Q_{Gi}^{\min} \leqslant Q_{Gi} \leqslant Q_{Gi}^{\max} \quad (5)$$

The voltage magnitude inequality constraints for the nodes are [21-24]:

$$V_i^{\min} \leqslant V_i \leqslant V_i^{\max} \tag{6}$$

Here $P_{Gi}^{\min}$ and $P_{Gi}^{\max}$ are the minimum and maximum active power outputs of generator $i$, $Q_{Gi}^{\min}$ and $Q_{Gi}^{\max}$ are the minimum and maximum reactive power outputs of generator $i$, $V_i^{\min}$ and $V_i^{\max}$ are the minimum and maximum voltage limits at node $i$.

## 3. Impact of Renewable Energy Integration on Power Flow Calculation

The introduction of renewable energy sources significantly impacts power flow calculations within power systems. Primarily, the intermittent and volatile nature of large-scale renewable energy generation introduces instability in output, resulting in dynamic changes and uncertainties in power distribution [25-28]. Traditional power flow calculations, which typically assume deterministic values for generation and load, are no longer suitable due to the variability of renewable sources. Moreover, the distributed integration of renewable energy alters the direction of power flows, necessitating the consideration of more nodes and a complex network topology, thus increasing the complexity and computational demand of calculations. Renewable energy primarily affects the network's topology, grid voltage, and power quality [29-31].

The entry of a substantial amount of new energy into the power network transforms the system from a single supply mode to various forms of power supply, altering the power distribution in the electricity market [32-35]. The diversity of distribution systems and their increasingly complex topologies make optimal power flow calculations more intricate. These complexities are reflected both at the interface between distributed generation (DG) and the main grid and in the complexity of flow distribution algorithms suitable for DG. Renewable energy sources impact node voltages within the power system, potentially leading to over-voltage and significant fluctuations primarily due to changes in line power [36-37]. Traditional distribution networks, which are unidirectional radial networks where power flows from the source towards the load, experience reduced transmission power on lines when integrated with distributed photovoltaics, potentially reversing the flow and causing voltage rises at various nodes. Operational variations caused by distributed generation can lead to voltage fluctuations or exceedances, especially when distributed resources are located near the end of the lines [38-41]. After integrating renewable sources into the grid, the system's power quality is affected due to the fluctuation and uncertainty of renewable sources, the frequent switching of large-scale distributed renewable resources, and interactions between renewable systems and voltage control devices [42-45].

Therefore, to accurately assess and predict the operational state of the grid, power flow calculations must consider the characteristics of renewable energy and employ more flexible and advanced methods to optimize network stability and economic efficiency [46].

## 4. Improved Optimal Power Flow Strategies

The core challenge of Optimal Power Flow (OPF) in power systems lies in optimizing control variables within a given system architecture, aiming to satisfy a range of grid constraints while ensuring an optimal power distribution that meets predefined objectives [47-49]. The integration of renewable energy sources on a large scale poses unprecedented challenges to conventional grid optimization and scheduling due to their significant variability and uncertainty [50-51]. To address these changes, current OPF strategies have been enhanced with advanced techniques such as stochastic optimization, robust optimization, distributed computing, data-driven real-time scheduling, and energy storage systems, providing flexible and reliable solutions for power systems with extensive renewable energy integration [53-56].

**4.1 Node Handling in Power Flow Calculation**

When discussing optimal power flow calculations involving wind farms, the core challenge focuses on how to effectively and appropriately handle the complex characteristics of wind power injection points (i.e., wind farm nodes) [57-62]. This process not only requires the accurate simulation of the uncertainty of wind power output but also ensures that the calculation model can flexibly respond to wind fluctuations to achieve optimal power flow distribution in the power system. Advanced algorithms and technical means must be used to replace traditional methods, thereby more accurately depicting the behavioral patterns of wind farm nodes in the power system and their impact on overall power flow [63-66]. Such strategic adjustments not only enhance the accuracy and reliability of the calculation results but also promote stable operation and optimized scheduling of the power system under high wind penetration [67].

In the power flow calculations after wind power grid integration, node processing methods for asynchronous wind turbines include: simplified power-voltage (P-V) equivalent models, PQ iterative refinement models [68], and more complex impedance-reactance (R-X) iterative models [69]. Going back to the initial stages of steady-state analysis of power systems, the academic community generally tended to abstract wind turbines as constant power (P-Q) type nodes for consideration [70] to simplify the analysis process. With the rapid development of wind technology and the expanding scale of grid integration, these traditional methods are gradually being replaced by more precise and comprehensive treatment strategies to better adapt to the intermittency and variability of wind output. The amount of reactive power absorbed by wind turbines is actually closely related to their slip rate and the output of active power, a characteristic that makes the simple classification of asynchronous generators as traditional P-Q nodes, P-V nodes, or balance nodes insufficient and potentially error-prone [71-76].

Therefore, more detailed and dynamic handling methods are needed to accurately reflect the actual behavioral characteristics of wind turbines in the power system, ensuring the accuracy and reliability of power flow calculation results. Reference [77] considers such wind turbine groups as a specific type of Q-V node based on the inherent properties of asynchronous wind generators. In the process of power system power flow calculations, it is only necessary to adjust related elements in the Jacobian matrix specifically, and the implementation of this strategy significantly optimizes the calculation process. Simulation data validation shows that using the Q-V model not only effectively reduces the number of necessary power flow calculations but also significantly reduces the overall calculation time, thus significantly enhancing computational efficiency and performance [78-80].

**4.2 Handling Uncertainty of Renewable Energy Output**

In addressing the OPF problem within power systems, traditional methodologies typically rest on a deterministic framework [81-83]. This often neglects the inherent stochastic elements, especially with the large-scale integration of variable renewable energy sources like wind [84-86]. Wind farms, whether connected through centralized or distributed strategies, exhibit substantial variability and unpredictability in their output due to wind speed fluctuations. This variability introduces complex and dynamic uncertainties that need careful consideration in OPF calculations [87-90].

Modern research on OPF in power systems bifurcates into two primary directions to handle these uncertainties. The first path involves constructing uncertainty models to thoroughly capture and quantify how wind variability impacts system states and decision-making [91-95]. The second approach enhances traditional deterministic models with stochastic handling techniques to simulate wind behavior more realistically, thus preserving computational efficiency while adapting to wind uncertainty [96-100]. This dual approach helps refine the theories and methods of power system OPF, aiming for greater accuracy, efficiency, and robustness.

Within the framework of uncertainty modeling, the generating capacities of wind and photovoltaics are treated as random variables. This has led to the development of advanced planning models like probability-based chance-constrained models, which set constraints that accommodate specific confidence levels to manage wind randomness [101-105]. Stochastic models employ probability distributions to detail wind output uncertainty, enhancing system analysis. Fuzzy models apply fuzzy numbers or sets for ambiguous or incomplete data, while interval analysis models handle wind output variations within predefined bounds, using interval arithmetic to assess system performance across different scenarios. These models form a comprehensive toolkit to tackle the randomness of wind energy effectively [106-110].

Several studies within the framework of uncertainty modeling in optimal power flow demonstrate diverse innovative approaches. Reference [111] presents a collaborative operation framework tailored for uncertain operational environments, incorporating an interval-based active power optimization flow model to adeptly handle system uncertainties. Reference [112] targets optimizing power dispatch strategies in wind-dominant systems using opportunity-constrained planning principles to effectively manage output fluctuations. Reference [113] utilizes Unscented Transformation to convert the complex probabilistic OPF problem into a deterministic form, streamlining the computational process while preserving crucial probabilistic features. Meanwhile, Reference [114] merges fuzzy and stochastic planning to develop a mixed chance-constrained economic dispatch model, capturing multiple uncertainties through advanced simulation techniques, albeit with increased computational demands. These methodologies collectively enhance the modeling and management of renewable integration into power systems.

For deterministic models, the scheduling period is finely segmented into discrete intervals, treating wind power output as a constant within each segment. This segmentation allows for precise mapping of wind speed variations and supports stable and reliable power system operations. However, this method might oversimplify the variability of wind power, potentially leading to conservative outcomes that ensure operational safety but may limit optimization potential [115-118].

Overall, these methodologies, from scenario-based to interval analysis, significantly enhance the ability to model and solve the intricate challenges of integrating high levels of renewable energy into power systems, supporting more effective and robust power system management.

**4.3 Optimal Power Flow Considering Energy Uncertainty**

Traditional OPF strategies, which are typically based on static loads and stable generation conditions, require more flexible optimization methods to handle the variability and uncertainty of new energy sources [119-121]. Approaches to manage this uncertainty include robust and stochastic optimization. Scenario optimization involves constructing numerous deterministic operating scenarios to analyze the uncertainty of new energy sources. Robust optimization considers the range of uncertain parameters and variables and system responses to find the best solution under the worst-case scenario [122]. Research has developed uncertainty sets using uncertainty and hypothesis testing methods to reduce the dimensionality of the problem, though the accuracy of solutions depends on empirically determined uncertainty sets, which can be limiting. The Chance Constrained Programming (CCP) method transforms new energy output constraints into probabilistic constraints, ensuring that new energy outputs meet operational requirements within a set confidence interval [123]. Gaussian assumptions for wind power outputs lead to reformulations of chance constraints into second-order cone inequalities for second-order cone programming. Efforts to establish line flow limits as chance constraints derive expressions for the cumulative distribution functions and their inverses of new energy sources, but numerical integration during this process can amplify errors, leading to inaccurate model solutions [124-126]. Research utilizing a Gaussian Mixture Model (GMM) for forecasting errors in wind and solar outputs has developed a Transient Stability Constrained OPF model incorporating rotating reserve chance constraints; this model uses sequence operations to deterministically transform chance constraints, simplifying the TSCOPF model into a mixed-integer linear programming problem that is easier to solve. Uncertainty modeling and analysis for OPF include robust and stochastic optimization approaches [127]. Further research is needed to develop more effective and flexible OPF calculation methods to meet real-time requirements, given the increasing complexity of power systems and the challenges of modeling uncertainty.

**4.4 Optimal Power Flow Considering Energy Storage Systems**

In the context of extensive renewable energy integration, developing optimal power flow with energy storage systems is crucial due to the intermittency and variability of renewable sources, which can lead to imbalances between power supply and demand [128]. As energy storage technology rapidly advances, systems like

Energy Storage Systems (ESS) are becoming practical. With the scaling of new energy sources, the proliferation of distributed generation sources, and increasing transmission corridor pressures, the development of new storage technologies becomes urgent [129-132]. Current large-scale storage systems utilize various methods such as lithium-ion batteries, hydrogen storage, and compressed air energy storage.

Addressing OPF issues with energy storage involves numerous constraints, as storage devices allow for bidirectional energy transfer within the power system, necessitating constraints over multiple time periods [133-135]. Existing research methods based on conventional OPF models perform optimal power flow calculations across many load profiles, complicating the problem and increasing the difficulty of obtaining solutions. A proposed model treats storage as a finite-horizon optimal control problem in OPF [136]. For specific cases of a single generator and load, the optimal generation plan may intersect the time-varying demand curve only once, implying that to charge the batteries, the optimal power flow initially produces more power than demanded, then less, and finally uses the batteries to supplement generation [137]. New algorithms demonstrate preliminary results of interactions between ESS and stochastic generators, motivated by studies on renewable energy generation and how to better utilize inherent oscillations in power output. ESS benefits in dynamic optimization exceed the actual scheduled amount of renewable energy in the system [138]. The potential adoption of electrified transportation could significantly increase the local distribution system's pressure. However, once the interface between the grid and vehicle power sources is extensively developed, the availability of distributed resources will also increase in the form of energy storage [139-141]. A proposed solution considers the entire system, taking into account storage devices, voltage, current, and power limits. The power network can be arbitrarily complex, and the proposed solver achieves a global optimum. Another strategy uses convex optimization-based relaxation to solve optimal control problems, illustrating the impact of different levels of storage using the IEEE benchmark system topology and time-invariant and demand-based cost functions [142]. The addition of energy storage and demand-based cost functions significantly reduces generation costs and flattens the generation curve [143-145].

**4.5 Distributed Optimal Power Flow**

With the large-scale integration of renewable energy in modern power systems, there are many controllable resources, and different control areas of the power system might differ in their control information [146-148]. Distributed optimal power flow is mostly used in DC flows to overcome the computational and communication bottlenecks brought by centralized optimization methods, better adapting to new power system configurations. Reference [149] discusses the use of Lagrangian relaxation for multi-zone distributed generation optimization scheduling, requiring real-time updates at the central node during iterative solutions to distribute Lagrange multipliers to various regions, which prevents full decentralization. Reference [150] has developed a new distributed voltage optimization algorithm based on the principle of auxiliary problems, which necessitates data exchange between regions. Reference [151] introduces a cost optimization method for DC systems based on the theory of consistency but does not incorporate the power system flow equations into the model. Reference [152] applies the Alternating Direction Method of Multipliers (ADMM) to develop a distributed optimal power flow method for DC distribution networks, which is suitable only for radial networks. Reference [153] uses a decomposition coordination approach, considering the impact of wind power integration into the grid by first dividing the system into several areas and then using global variables to handle boundary node issues, minimizing information transfer by limiting it to adjacent sub-regions to enable distributed optimal power flow calculation. Reference [154] initially applies a second-order cone relaxation to the DC system's optimal power flow model to create a second-order cone optimal power flow model and then constructs a distributed optimal power flow algorithm based on ADMM on this model, subsequently modifying the algorithm to a fully distributed method by removing the consistency variables associated with the neighboring regional boundary nodes.

**4.6 Optimal Power Flow Based on Linearization**

Optimal power flow is typically a non-linear, non-convex optimization problem that requires relaxation and solution through various optimization methods [155-158]. However, the widespread application of linearization to flow equations due to its effective control capabilities has garnered widespread attention. The integration of renewable energy complicates the network structure and flow characteristics, making non-linear approaches unsuitable for online applications. Reference [159] builds a model for optimizing the capacity to integrate new energy based on uncertain factors, along with a multi-objective optimization prediction model for high penetration of renewable energies into the power network. Reference [160] gives a method for calculating the unbalanced probabilistic trends in active distribution networks based on linearized forward-back substitution principles. Reference [161] aims to use a new information-physical fusion-based power system linearization algorithm for three-phase optimal power flow distribution, achieving higher accuracy and maintaining it under large loads. Reference [162] proposes improvements in linearized flow algorithm accuracy and the design of enhanced DC flow models for voltage and reactive power. Reference [163] studies and linearizes the mathematical models of bipolar DC distribution systems to provide suitable flow calculations for multiple voltage levels.

**4.7 Data-Driven Optimal Power Flow**

As power system flow models grow in complexity, traditional model-based optimal power flow calculations reveal their limitations, prompting a shift toward real-time, efficient analysis using data-driven approaches that have gained significant attention in recent years [164-166]. Currently, data-driven optimal power flow methods mainly fall into two categories: supervised learning and reinforcement learning (RL). These methods facilitate effective problem-solving by mapping relationships between optimal power flow inputs and outputs [167-168]. Reference [169] suggests a method for quantifying the reliability of optimal power flow results calculated through deep neural networks by evaluating mapping errors theoretically and updating parameters to establish credible deep neural network-based calculations. Reference [170] first predicts optimal power flow voltages using deep neural networks and then determines the remaining optimal power flow results by solving linear equations. Reference [171] introduces an optimal power flow calculation technique based on deep neural networks. Reference [172] incorporates physical information of power systems into the training of deep neural networks, achieving efficient and accurate learning outcomes. Reference [173] uses unit active power outputs to represent phase angles in the output mapping, effectively reducing the mapping dimensionality and decreasing the size of the deep neural network model and the amount of training data required. It employs uniform sampling to avoid common overfitting issues in generic deep neural network methods. Reference [174] investigates methods for identifying critical decision parameters in deep networks for optimal power flow, combining correlation analysis and clustering to explore the match between inputs and outputs and constructing a segmented feature library to simplify the training process. Reference [175] designs a deep neural network training strategy based on sensitivity to load levels derived from the partial derivatives of optimal power flow calculation results, aimed at enhancing the constraint satisfaction capability of deep neural networks. Supervised learning enhances the efficiency of optimal power flow calculations, but the "black box" nature of neural networks means their interpretability needs further exploration [176-180]. Learning the process of solving optimal power flow problems through the interaction between intelligent agents and the environment encapsulates the core idea behind reinforcement learning-based optimal computation methods. Reference [181] describes a real-time optimal power flow calculation method based on reinforcement learning techniques. Reference [182] discusses the application of deep reinforcement learning in the field of power system voltage control.

Data-driven optimal power flow methods do not require the construction of precise and complex mathematical models [183-185]. Instead, they mine relationships through massive data sets, but the outcomes highly depend on the quantity and quality of data needed for training, lacking interpretability and generalizability, and are not yet broadly applicable in practical engineering [186-187]. Overall, the application of data-driven technologies enables high-accuracy and rapid flow analyses, providing a robust technical foundation for swiftly resolving and assessing optimal power flow under constrained conditions [188].

**5 Conclusion**

This paper explores optimal power flow strategies in new power systems that incorporate renewable energies, analyzing the main technical challenges posed by the variability and uncertainty of renewable resources, summarizing existing improved optimal power flow strategies, and suggesting possible future trends based on current research status and technological prospects.

1) With the increase of renewable energy in power systems, traditional operations and planning are challenged. The traditional optimal power flow problem, based on stable loads and generators, faces issues like severe voltage fluctuations and a decline in power quality as renewable energy's variability and uncertainty are introduced.

2) In response to challenges from renewable energies, researchers have proposed new improved methods and algorithms. Traditional flow calculations, based on physical mechanisms, are precise but slow, failing to meet real-time computation requirements in large-scale, complex power systems.

3) Considering the uncertainty of renewable energy sources, optimal power flow analysis can reflect the impact of various random factors on system operation. However, achieving this requires constructing uncertain flow models for power systems, where model accuracy directly affects computation precision. Considering energy storage optimization in flow calculations while accounting for the efficiency of storing and releasing energy, and implementing segmental corrections and relaxations of residual energy constraints, enables efficient energy use under normal storage and release demands; data-driven methods significantly enhance flow calculation speed, making high-accuracy rapid flow analysis feasible. Linearized optimal power flow offers unique advantages in addressing OPF issues in meshed distribution networks, such as ensuring convergence, high computational efficiency, and easy access to nodal marginal price information.

With the national dual carbon goals and increasing demand for renewable energy, the stable operation and optimized control of power systems face numerous difficulties and challenges. In the coming years, as technology advances and energy policies progress, power systems will evolve toward more intelligent and flexible management. Firstly, to more efficiently balance the intermittency and volatility of renewable sources, combining big data analysis and artificial intelligence algorithms will enable real-time optimal power flow, reducing losses and enhancing energy utilization efficiency. The volatility and uncertainty of renewable energy have driven traditional optimal power flow strategies toward more flexible, intelligent directions. By introducing advanced algorithms and technologies, while ensuring the stability of power systems, the utilization of renewable energy can be maximized to achieve an economical and environmentally friendly operation of power systems.

**References**

[1]National Energy Administration. The national total power generation capacity has more than 3 billion kw [EB/OL]. (2024-05-31) [2024-05-31]. http://www.nea.gov.cn/2024-05/31/c_1310776691.htm.

[2]Pan Xiang, Zhao Tianyu, Chen Minghua. DeepOPF deep neural network for DC optimal power flow[C]//2019 IEEE International Conference on Communication, Control, and Computing Technologies for Smart Grids, Beijing, China, 2019: 1-6.

[3]Liang Junwen, Lin Shunjiang, Liu Mingbo. Distributed reactive power optimization control method for active distribution network [J]. Power Grid Technology, 2018, 42(1):230-237.

[4]Wang Fangzong, Wang Zhaofeng. Optimal reconstruction method of distribution Network with Distributed Power based on mixed integer second-order cone programming [J]. Power System Protection and Control, 2016, 44(24):24-30.

[5]Gao Hongjun, Liu Junyong, Shen Xiaodong, et al. Research on optimal Power Flow of Active Distribution Network and its Application [J]. Proceedings of the CSEE, 2017, 37(6):1634-1645.

[6]Zhang Boming, Chen Shousun. Advanced Power network analysis [M]. Beijing: Tsinghua University Press, 1996.

[7]Sun D, Ashley B, Brewer B, et al. Optimal power flow by Newton approach[J]. IEEE Transactions on Power Apparatus and Systems, 1984, PAS-103(10): 2864- 2880.

[8]Jabr R A, Coonick A H, Cory B J. A primal-dual interior point method for optimal power flow dispatching[J]. IEEE Transactions on Power Systems, 2002, 17(3): 654-662.

[9]Zhang Wei, Wei Zhinong, Liu Yujuan. Chaos optimal algorithm based power flow calculation involving wind farms[J]. Electric Power, 2011, 44(10): 25-28.

[10]Gu Chenghong，Ai Qian. Optimal power flow calculation based on the improved interior method for a system integrated with wind farms[J].Electric Power, 2007, 40(1): 89-93.

[11]Chen Jinfu, Chen Haiyan, Duan Xianzhong. Multi-period dynamic optimal power flow in wind power integrated system[J]. Proceedings of the CSEE, 2006, 26(3): 31-35.

[12]Fan Rongqi, Chen Jin Fu, Duan Xianzhong，et al. Impact of wind speed correlation on probabilistic load flow[J].Automation of Electric Power Systems, 2011, 35(4):18-22+76.

[13]Deng Wei, Li Xinran, Xu Zhenghua，et al. Calculation of probabilistic load flow considering wind speed correlation and analysis on influence of wind speed correlation[J]. Power System Technology, 2012, 36(4): 45-50.

[14]Sun Guoqiang, Li Yichi，Xiang Yupeng, et al. Dynamic stochastic optimal power flow of wind integrated power system considering temporal and spatial correlation of wind speed[J]. Proceedings of the CSEE, 2015, 35(17): 4308-4317.

[15]Luo Jiayong, Bao Haibo, Guo Xiaoxuan. Calculation of probabilistic optimal power flow considering wind speed correlation of wind farm[J]. Guangxi Electric Power, 2014, 37(1): 13-17.

[16]Liao Yingchen, Gan Deqiang, Chen Xingying, et al. Fuzzy optimal power flow analysis considering indeterminacy of distributed generation for urban power grid[J]. Electric Power Automation Equipment, 2012, 32(9): 35-39

[17]Li Y, Wang J, Zhao D, et al. A two-stage approach for combined heat and power economic emission dispatch: Combining multi-objective optimization with integrated decision making[J]. Energy, 2018, 162: 237-254.

[18]Lin Haiming, Liu Tianqi, Li Xingyuan. Optimal active power flow considering uncertainties of wind power output and load[J].Power System Technology, 2013, 37(6): 1584-1589.

[19]Bao Haibo, Wei Hua. Probabilistic optimal power flow computation in power systems including large-scale wind farms based on unscented transformation[J]. Automation of Electric Power Systems, 2014, 38(12): 46-53.

[20]Chen Honggui, Li Zhihuan, Chen Jinfu, et al. SFL algorithm based dynamic optimal power flow in wind power integrated system[J]. Automation of Electric Power Systems, 2009, 33(4):25-30.

[21]W. Huang, X. Pan, M. Chen, and S. H. Low. Deep OPF-V: Solving AC-OPF problems efficiently[J]. IEEE Transactions on Power Systems, 2022, 37(1): 800-803.

[22]M. Chatzos, F. Fioretto, T. W. Mak, and P. Van Hentenryck. High-fidelity machine learning approximations of large-scale optimal power flow[J]. arXiv, 2020: arXiv:2006.16356.

[23]Z. Yan and Y. Xu, "Real-time optimal power flow: A lagrangian based deep reinforcement learning approach[J]. IEEE Transactions on Power Systems, 2020, 35(4): 3270-3273.

[24]S. Wang et al. A data-driven multi-agent autonomous voltage control framework using deep reinforcement learning[J]. IEEE Transactions on Power Systems, 2020, 35(6): 4644-4654.

[25]Wang Yongtao, Wang Linhong, Luo Songtao. The application of novel hybrid particle swarm algorithm for optimal power flow calculation of power systems with considering wind power integration[J]. Renewable Energy Resources, 2013, 31(6): 34-37.

[26]Li Y, Li Y. Two-step many-objective optimal power flow based on knee point-driven evolutionary algorithm[J]. Processes, 2018, 6(12): 250.

[27]ZHUO Z Y, ZHANG N, YANG J W, et al. Transmission expansion planning test system for ac/dc hybrid grid with high variable renewable energy penetration[J]. IEEE Transactions on Power Systems, 2020, 35(4):2597-2608.

[28]Ju Yuntao, Yang Mingyou, Wu Wenchuan. Data Physics Fusion Drive Linearization Method for three-phase power flow optimization in distribution network [J]. Automation of Electric power Systems, 2022, 46(13):43-52.

[29]DI FAZIO A R，RISI C，RUSSO M，et al. Distributed voltage optimization based on the auxiliary problem prin-ciple in active distribution systems [C]//2020 55th International Universities Power Engineering Conference(UPEC). September 1-4，2020. Turin，Italy：IEEE, 2020.1-6.

[30]Dong Mi, Zhang Xinlu, Yang Jian, et al. Coordinated optimal control strategy for economic operation of multi-source DC microgrid in island mode [J]. New Technology of Electroengineering and Electric Energy, 2019, 38(5):51-58.

[31]Han Yuxin, Chen Laijun, Wang Zhaojian, et al. Distributed optimal Power Flow of DC Distribution Network based on adaptive Step Size ADMM [J]. Transactions of China Electrotechnical Society, 2017, 32(11):26-37.

[32]Huy T H B, Nguyen T P, Nor N M, et al. Performance improvement of multiobjective optimal power flow-based renewable energy sources using intelligent algorithm[J]. IEEE Access, 2022, 10: 48379-48404.

[33]Shaheen M A M, Hasanien H M, Al-Durra A. Solving of optimal power flow problem including renewable energy resources using HEAP optimization algorithm[J]. IEEE Access, 2021, 9: 35846-35863.

[34]Wang Z, Anderson C L. A progressive period optimal power flow for systems with high penetration of variable renewable energy sources[J]. Energies, 2021, 14(10): 2815.

[35]Li Y, Feng B, Wang B, Sun S. Joint planning of distributed generations and energy storage in active distribution networks: A Bi-Level programming approach[J]. Energy, 2022, 245: 123226.

[36]GAYME D, TOPCU U. Optimal power flow with large-scale storage integration[J]. IEEE Trans on Power Systems, 2013, 28(2): 709-717.

[37]YINGVIVATANAPONG C，LEE W J，LIU E. Multi-area power generation dispatch in competitive markets[J]. IEEE Transactions on Power Systems，2008，23(1):196-203.

[38]Li Guoqing, Hui Xinxin, Sun Fujun, et al. Distributed DC-OPF calculation of wind farm system based on ADMM [J]. Journal of Solar Energy, 2019, 40(8):2178-2186.

[39]Li Y, Yang Z. Application of EOS-ELM with binary Jaya-based feature selection to real-time transient stability assessment using PMU data[J]. IEEE Access, 2017, 5: 23092-23101.

[40]Wu Kui-Hua, Li Wen-Sheng, Wu Zhao-yun, et al. Distributed optimal power flow scheduling for multi-energy complementary microgrids based on alternating direction multiplier [J]. Power Grids and Clean Energy, 2022, 38 (11) :35-44.

[41]Zhu Shu, Liu Kaipei, Qin Liang, et al. Transient Stability Analysis of Electronic Power Systems [J]. Proceedings of the CSEE, 2017,37 (14) : 3948-3962, 4273.

[42]Wang Qun, Dong Wenlue, Yang Li. Wind power/photovoltaic classical scene set generation algorithm based on Wasserstein distance and improved K-medoids clustering [J]. Proceedings of the CSEE, 2015, 35(11):2654-2661.

[43]Li Y, Li Y, Li G. Security-constrained multi-objective optimal power flow for a hybrid AC/VSC-MTDC system with lasso-based contingency filtering[J]. IEEE Access, 2019, 8: 6801-6811.

[44]DUAN C，JIANG L，FANG W，et al. Data-driven affinely adjustable distributionally robust unit commitment[J]. IEEE Transactions on Power Systems，2018，33(2): 1385-1398.

[45]BIENSTOCK D，CHERTKOV M，HARNETT S. Chance constrained optimal power flow: risk-aware network control under uncertainty[J]. SIAM REVIEW，2014，56（3）：461-495.

[46]Li Y, Li Y, Li G. Optimal power flow for AC/DC system based on cooperative multi-objective particle swarm optimization[J]. Automation of Electric Power Systems, 2019, 43: 94-100.

[47]Wu Xinzhang, Guo Suhang, Dai Wei, et al. Probabilistic optimal Power flow deep learning method for transmission network based on feature dimensionality reduction and segmentation [J]. Electric Power Automation Equipment, 2023, 43(8): 174-180.

[48]Zou Mingjun，Chen Jinghua，Tang Junjie. A review of the research on the optimal power flow of wind farms and its key technologies[J]. Journal of Guangdong University of Technology, 2018,35(02):63-68+94.

[49]ZHU Rujie，WEI Hua，BAI Xiaoqing. Distributionally robust optimization of multi-energy dynamic optimal power flow[J].Proceedings of the CSEE，2020，40（11）：3489-3497.

[50]Huang Minghao，Chen Yifeng，Dong Shufeng. Low-carbon optimal currents for transmission and distribution cooperation based on Anderson acceleration[J]. Power System Technology，2023，47（8）：3132-3144.

[51]Li Y, Han M, Yang Z, et al. Coordinating flexible demand response and renewable uncertainties for scheduling of community integrated energy systems with an electric vehicle charging station: A bi-level approach[J]. IEEE Transactions on Sustainable Energy, 2021, 12(4): 2321-2331.

[52]Peng Yue, Xiong Wei, Yuan Xufeng, et al. Research on optimal Power Flow of Active Distribution Network based on mixed integer second-order cone Programming[J]. Electrical Measurement & Instrumentation, 2023,60 (5) : 139-144.

[53]Li Y, He S, Li Y, et al. Renewable energy absorption oriented many-objective probabilistic optimal power flow[J]. IEEE Transactions on Network Science and Engineering (Early Access), 2023. DOI: 10.1109/TNSE.2023.3290147

[54]Vrakopoulou M , Margellos K , Lygeros J , et al. A probabilistic framework for security constrained reserve scheduling of networks with wind power generation[C]// Energy Conference & Exhibition, IEEE, 2012.

[55]Phan D , Kalagnanam J . Distributed methods for solving the security-constrained optimal power flow problem[C]// IEEE. IEEE, 2012.

[56]Wang Z , Yan X , Zhao Y D , et al. An effcient approach for robust SCOPF considering load and renewable power uncertainties[C]// Power Systems Computation Conference. IEEE, 2016.

[57]Q. W,J. D M,T.Z, et al. A Computational Strategy to Solve Preventive Risk-Based Security-Constrained OPF[J]. lEEE Transactions on Power Systems.2013,28(2):1666-1675.

[58]Yu Zhen, Shen Shen, Liu Feng, et al. Economic scheduling with transient stability constraints considering wind power Uncertainty [J]. Proceedings of the CSEE, 2020, 40(20):7270-7282.

[59]Chen Ping, Dang Xi, Liu Longcheng, et al. Research on transient stability constrained optimal Power flow model considering wind and landscape Uncertainty [J]. Smart Power, 2024, 52(03):17-24.

[60]WU H，YUAN Y，ZHU J P，et al. Potential assessment of spatial correlation to improve maximum distributed PV hosting capacity of distribution networks[J]. Journal of Modern Power Systems and Clean Energy，2021，9(4):800-810.

[61]Li Y, Zhang M, Chen C. A deep-learning intelligent system incorporating data augmentation for short-term voltage stability assessment of power systems[J]. Applied Energy, 2022, 308: 118347.

[62]XIE H L, LI B, HEYMAN C, et al. Subsynchronous resonance characteristics in presence of doubly-fed induction generator and series compensation and mitigation of subsynchronous resonance by proper control of series capacitor[J]. IET Renewable Power Generation, 2014,8(4): 411-421.

[63]LIU Y C, RAZA A, ROUZBEHI K, et al. Dynamic resonance analysis and oscillation damping of multiterminal DC grids[J]. IEEE Access, 2017, 5: 16974-16984.

[64]Stott B, Hobson E. Power System Security Control Calculations Using Linear Programming, Part II[J]. lEEE Transactions on Power Apparatus and Systems, 1978, PAS.97(5):1721-1731.

[65]Burchett, R. C, Happ, et al. Large Scale Security Dispatching: An Exact Model[J]. Power Engineering Review, EEE, 1983.

[66]Li Y, Bu F, Li Y, et al. Optimal scheduling of island integrated energy systems considering multi-uncertainties and hydrothermal simultaneous transmission: A deep reinforcement learning approach[J]. Applied Energy, 2023, 333: 120540.

[67]Hmida J B, Chambers T, Lee J. Solving constrained optimal power flow with renewables using hybrid modified imperialist competitive algorithm and sequential quadratic programming[J]. Electric Power Systems Research, 2019, 177: 105989.

[68]Liu F, Li Y, Li B, et al. Bitcoin transaction strategy construction based on deep reinforcement learning[J]. Applied Soft Computing, 2021, 113: 107952.

[69]Li Y, Cao J, Xu Y, et al. Deep learning based on Transformer architecture for power system short-term voltage stability assessment with class imbalance[J]. Renewable and Sustainable Energy Reviews, 2024, 189: 113913.

[70]Hou Y, Jia S, Lun X, et al. Deep feature mining via attention-based BiLSTM-GCN for human motor imagery recognition[J]. arXiv preprint arXiv:2005.00777, 2020.

[71]Cui B , Sun X A . A New Voltage Stability-Constrained Optimal Power Flow Model: Sufficient Condition, SOCP Representation, and Relaxation[J]. lEEE Transactions on Power Systems, 2017:1-1.

[72]Lubin M , Dvorkin Y , Roald L . Chance Constraints for Improving the Security of AC Optimal Power Flow[J]. IEEE Transactions on Power Systems, 2019, 34(3):1908-1917.

[73]Roald L , Oldewurtel F , Parys B V , et al. Security Constrained Optimal Power Flow with Distributionally Robust Chance Constraints[J]. Mathematics, 2015.

[74]Liang F , Li S , He G , et al. A New Algorithm for Quadratic Programming with Interval-Valued Bound Constraints[C]// international Conference on Natural Computation. IEEE.2008.

[75]Li Y, Wang R, Li Y, et al. Wind power forecasting considering data privacy protection: A federated deep reinforcement learning approach[J]. Applied Energy, 2023, 329: 120291.

[76]Samakpong T, Ongsakul W, Madhu Manjiparambil N. Optimal power flow incorporating renewable uncertainty related opportunity costs[J]. Computational Intelligence, 2022, 38(3): 1057-1082.

[77]Hasanien H M, Alsaleh I, Ullah Z, et al. Probabilistic optimal power flow in power systems with renewable energy integration using enhanced walrus optimization algorithm[J]. Ain Shams Engineering Journal, 2024, 15(3): 102663.

[78]Li Y, Wang C, Li G, et al. Improving operational flexibility of integrated energy system with uncertain renewable generations considering thermal inertia of buildings[J]. Energy Conversion and Management, 2020, 207: 112526.

[79]Nie Y, Du Z, Li J. AC–DC optimal reactive power flow model via predictor–corrector primal-dual interior-point method[J]. IET Generation, Transmission & Distribution, 2013, 7(4): 382-390.

[80]Yuan Y, Kubokawa J, Sasaki H. A solution of optimal power flow with multicontingency transient stability constraints[J]. IEEE Transactions on Power Systems, 2003, 18(3): 1094-1102.

[81]Mo N, Zou ZY, Chan KW, Pong TYG. Transient stability constrained optimal power flow using particle swarm optimisation[J]. IET Generation, Transmission & Distribution, 2007, 1(3): 450-459.

[82]Li Y, Wang B, Yang Z, et al. Optimal scheduling of integrated demand response-enabled community-integrated energy systems in uncertain environments[J]. IEEE Transactions on Industry Applications, 2021, 58(2): 2640-2651.

[83]Zhao Fei, Fan Xuejun, Li Yalou, et al. Power flow calculation and optimal power flow model of network distribution Network based on iterative implicit linearization [J]. Proceedings of the CSEE, 2024, 44(06):2197-2208.

[84]Ran Qing-Yue, Lin Wei, Yang Zhi-Fang, et al. Optimal power flow calculation method based on trusted deep neural network [J/OL]. Transactions of China Electrotechnical Society, 2024, 1-14.

[85]W. Huang, X. Pan, M. Chen, and S. H. Low. Deep OPF-V: Solving AC-OPF problems efficiently[J]. IEEE Transactions on Power Systems, 2022, 37(1): 800-803.

[86]Li Y, Wang C, Li G, et al. Optimal scheduling of integrated demand response-enabled integrated energy systems with uncertain renewable generations: A Stackelberg game approach[J]. Energy Conversion and Management, 2021, 235: 113996.

[87]Niknam T, Narimani MR, Aghaei J, Azizipanah-Abarghooee R. Improved particle swarm optimisation for multi-objective optimal power flow considering the cost, loss, emission and voltage stability index[J]. IET Generation, Transmission & Distribution, 2012, 6(6): 515-527.

[88]Street A, Moreira A, Arroyo JM. Energy and reserve scheduling under a joint generation and transmission security criterion: An adjustable robust optimization approach[J]. IEEE Transactions on Power Systems, 2014, 29(1): 3-14.

[89]Li Y, Li Z, Chen L. Dynamic state estimation of generators under cyber attacks[J]. IEEE Access, 2019, 7: 125253-125267.

[90]Feng W, Tuan LA, Tjernberg LB, Mannikoff A, Bergman A. A new approach for benefit evaluation of multiterminal VSC–HVDC using a proposed mixed AC/DC optimal power flow[J]. IEEE Transactions on Power Delivery, 2014, 29(1): 432-443.

[91]Li Y, Wei X, Li Y, et al. Detection of false data injection attacks in smart grid: A secure federated deep learning approach[J]. IEEE Transactions on Smart Grid, 2022, 13(6): 4862-4872.

[92]Biswas P P, Suganthan P N, Amaratunga G A J. Optimal power flow solutions incorporating stochastic wind and solar power[J]. Energy conversion and management, 2017, 148: 1194-1207.

[93]Chen S Y, Chang C H. Optimal power flows control for home energy management with renewable energy and energy storage systems[J]. IEEE Transactions on Energy Conversion, 2022, 38(1): 218-229.

[94]Li Y, Gu X. Power system transient stability assessment based on online sequential extreme learning machine[C]//2013 IEEE PES Asia-Pacific Power and Energy Engineering Conference (APPEEC). IEEE, 2013: 1-4.

[95]Khaled U, Eltamaly A M, Beroual A. Optimal power flow using particle swarm optimization of renewable hybrid distributed generation[J]. Energies, 2017, 10(7): 1013.

[96]Cao D, Hu W, Xu X, et al. Deep reinforcement learning based approach for optimal power flow of distribution networks embedded with renewable energy and storage devices[J]. Journal of Modern Power Systems and Clean Energy, 2021, 9(5): 1101-1110.

[97]Li Y, Li G, Wang Z. Rule extraction based on extreme learning machine and an improved ant-miner algorithm for transient stability assessment[J]. PloS one, 2015, 10(6): e0130814.

[98]Singh M, Kekatos V, Giannakis G. Learning to solve the AC-OPF using sensitivity-informed deep neural networks[C]//2022 IEEE Power & Energy Society General Meeting (PESGM), Denver, CO, USA, 2022: 1.

[99]Nusair K, Alhmoud L. Application of equilibrium optimizer algorithm for optimal power flow with high penetration of renewable energy[J]. Energies, 2020, 13(22): 6066.

[100]GOLSHANNAVAZ S, AMINIFAR F, NAZARPOUR D. Application of UPFC to enhancing oscillatory response of series-compensated wind farm integrations[J]. IEEE Transactions on Smart Grid, 2014,5(4): 1961-1968.

[101]Yang Z , Zhong H , Bose A , et al. Optimal Power Flow in AC-DC Grids with Discrete Control Devices[J]. IEEE Transactions on Power Systems, 2017:1-1.

[102]Lejeune M A,Dehghanian P. Optimal Power Flow Models With Probabilistic Guarantees: A Boolean Approach[J]. IEEE Transactions on Power Systems, 2020, 35(6): 4932 - 4935.

[103]Li Y, Li J, Wang Y. Privacy-preserving spatiotemporal scenario generation of renewable energies: A federated deep generative learning approach[J]. IEEE Transactions on Industrial Informatics, 2021, 18(4): 2310-2320.

[104]Zhao Shuben, Zhang Fusheng. Solution of optimal power flow based on differential evolution and its modified algorithm[J].Power System Technology, 2010, 34(8): 123-128.

[105]Khan B, Singh P. Optimal power flow techniques under characterization of conventional and renewable energy sources: A comprehensive analysis[J]. Journal of Engineering, 2017, 2017(1): 9539506.

[106]Reddy S S. Optimal power flow with renewable energy resources including storage[J]. Electrical Engineering, 2017, 99: 685-695.

[107]Li Y, Han M, Shahidehpour M, et al. Data-driven distributionally robust scheduling of community integrated energy systems with uncertain renewable generations considering integrated demand response[J]. Applied Energy, 2023, 335: 120749.

[108]Karthik N, Parvathy A K, Arul R, et al. Multi-objective optimal power flow using a new heuristic optimization algorithm with the incorporation of renewable energy sources[J]. International Journal of Energy and Environmental Engineering, 2021, 12: 641-678.

[109]Adhikari A, Jurado F, Naetiladdanon S, et al. Stochastic optimal power flow analysis of power system with renewable energy sources using Adaptive Lightning Attachment Procedure Optimizer[J]. International Journal of Electrical Power & Energy Systems, 2023, 153: 109314.

[110]Zhou Niancheng, Zhang Yu, Liao Jianquan, et al. Linearized Power flow Calculation of Multi-voltage class bipolar DC Distribution Network with DC Transformer Control [J]. Proceedings of the CSEE, 2022, 42(6):2070-2084.

[111]M. Chatzos, F. Fioretto, T. W. Mak, and P. Van Hentenryck. High-fidelity machine learning approximations of large-scale optimal power flow[J]. 2020:16356.

[112]Z. Yan and Y. Xu, "Real-time optimal power flow: A lagrangian based deep reinforcement learning approach[J]. IEEE Transactions on Power Systems, 2020, 35(4): 3270-3273.

[113]Li Y, Li Y, Li G. Two-stage multi-objective OPF for AC/DC grids with VSC-HVDC: Incorporating decisions analysis into optimization process[J]. Energy, 2018, 147: 286-296.
17

[114]LEVRON Y, GUERRERO J M, BECK Y. Optimal power flow in microgrids with energy storage[J]. IEEE Trans on Power Systems, 2013, 28(3): 3226-3234.

[115]Hashish M S, Hasanien H M, Ullah Z, et al. Giant trevally optimization approach for probabilistic optimal power flow of power systems including renewable energy systems uncertainty[J]. Sustainability, 2023, 15(18): 13283.

[116]ALISMAIL F,XIONG P,SINGH C. Optimal wind farm allocation in multi-area power systems using distributionally robust optimization approach[J]. IEEE Transactions on Power Systems,2018,33(1): 536-544.

[117]Huang Minghao, Chen Yifeng, Dong Shufeng. Collaborative low-carbon optimal power flow for transmission and distribution based on Anderson acceleration [J]. Power Grid Technology, 2023, 47(8):3132-3144.

[118]WANG Z,SHEN C,LIU F,et al. Chance constrained economic dispatch with non-Gaussian correlated wind power uncertainty[J]. IEEE Transactions on Power Systems, 2017, 32(6): 4880-4893.

[119]ABAD M S,MA J,ZHANG D W,et al. Probabilistic assessment of hosting capacity in radial distribution systems［J］. IEEE Transactions on Sustainable Energy, 2018, 9(4):1935- 1947.

[120]Zhu Junpeng, Huang Yong, Ma Liang, et al. Evaluation of consumption capacity of distributed generation based on uncertain optimal power flow [J]. Automation of Electric Power Systems, 2022, 46(14):46-54.

[121]Ye Qing-Quan, Wu Ming-Qi, Wu Xu-guang, et al. Fully distributed optimal Power Flow Algorithm for Multi-region DC System based on ADMM [J]. Zhejiang Electric Power, 2024, 43(02):13-24.

[122]Zhu Haohao, Zhu Jizhong, Li Shenglin, et al. Power Flow Calculation and Voltage Analysis of Fractional Power Systems [J]. Electric Machines and Control, 2022, 26 (4) : 38-46.

[123]Li Y, Yang Z, Li G, et al. Optimal scheduling of an isolated microgrid with battery storage considering load and renewable generation uncertainties[J]. IEEE Transactions on Industrial Electronics, 2018, 66(2): 1565-1575.

[124]CHANDY K M, LOW S H, TOPCU U, et al. A simple optimal power flow model with energy storage[C] //Decision and Control (CDC), 2010 49th IEEE Conference on, IEEE, 2010: 1051-1057.

[125]LAMADRID A J, MOUNT T D, SHOEMAKER C. Dynamic optimization for the management of stochastic generation and storage[C] // Transmission and Distribution Conference and Exposition, Latin America (T&D-LA), 2010 IEEE/PES, IEEE, 2010: 860-866.

[126]S. Wang et al. A data-driven multi-agent autonomous voltage control framework using deep reinforcement learning[J]. IEEE Transactions on Power Systems, 2020, 35(6): 4644-4654.

[127]Maheshwari A, Sood Y R, Jaiswal S. Investigation of optimal power flow solution techniques considering stochastic renewable energy sources: Review and analysis[J]. Wind Engineering, 2023, 47(2): 464-490.

[128]Li Y, et al. Bi-level programming of distributed generation in active distribution network considering integration influence of energy storage system[J]. Acta Energiae Sol. Sin, 2017, 38: 3311-3318.

[129]Alasali F, Nusair K, Obeidat A M, et al. An analysis of optimal power flow strategies for a power network incorporating stochastic renewable energy resources[J]. International Transactions on Electrical Energy Systems, 2021, 31(11): e13060.

[130]Syed M S, Chintalapudi S V, Sirigiri S. Optimal power flow solution in the presence of renewable energy sources[J]. Iranian Journal of Science and Technology, Transactions of Electrical Engineering, 2021, 45: 61-79.

[131]Nguyen T T, Nguyen H D, Duong M Q. Optimal power flow solutions for power system considering electric market and renewable energy[J]. Applied Sciences, 2023, 13(5): 3330.

[132]Wang Jing, Chen Junyu, Lan Kai. PSO optimal power flow algorithm for a microgrid based on spot power prices[J].Power System Protection and Control, 2013, 41(16): 34-40.

[133]Nusair K, Alhmoud L. Application of equilibrium optimizer algorithm for optimal power flow with high penetration of renewable energy[J]. Energies, 2020, 13(22): 6066.

[134]Mohamed A A, Kamel S, Hassan M H, et al. Optimal Power Flow Incorporating Renewable Energy Sources and FACTS Devices: A Chaos Game Optimization Approach[J]. IEEE Access, 2024.

[135]Guo Z, Wei W, Chen L, et al. Parametric distribution optimal power flow with variable renewable generation[J]. IEEE Transactions on Power Systems, 2021, 37(3): 1831-1841.

[136]Hassan M H, Kamel S, Alateeq A, et al. Optimal power flow analysis with renewable energy resource uncertainty: a hybrid AEO-CGO approach[J]. IEEE Access, 2023, 11: 122926 - 122961.

[137]Rambabu M, Nagesh Kumar G V, Sivanagaraju S. Optimal power flow of integrated renewable energy system using a thyristor controlled SeriesCompensator and a grey-wolf algorithm[J]. Energies, 2019, 12(11): 2215.

[138]Shargh S, Mohammadi-Ivatloo B, Seyedi H, et al. Probabilistic multi-objective optimal power flow considering correlated wind power and load uncertainties[J]. Renewable Energy, 2016, 94: 10-21.

[139]Abdullah M, Javaid N, Khan I U, et al. Optimal power flow with uncertain renewable energy sources using flower pollination algorithm[C]//Advanced Information Networking and Applications: Proceedings of the 33rd International Conference on Advanced Information Networking and Applications (AINA-2019) 33. Springer International Publishing, 2020: 95-107.

[140]Saha A, Bhattacharya A, Das P, et al. A novel approach towards uncertainty modeling in multiobjective optimal power flow with renewable integration[J]. International Transactions on Electrical Energy Systems, 2019, 29(12): e12136.

[141]Hassan M H, Mohamed E M, Kamel S, et al. Stochastic Optimal Power Flow Integrating with Renewable Energy Resources and V2G Uncertainty Considering Time-Varying Demand: Hybrid GTO-MRFO Algorithm[J]. IEEE Access, 2024, 12: 97893 - 97923.

[142]Shi Z, Yu T, Zhao Q, et al. Comparison of algorithms for an electronic nose in identifying liquors[J]. Journal of Bionic Engineering, 2008, 5(3): 253-257.

[143]Bo X, Chen X, Li H, et al. Modeling method for the coupling relations of microgrid cyber-physical systems driven by hybrid spatiotemporal events[J]. IEEE Access, 2021, 9: 19619-19631.

[144]Pan J, Dong A, Fan J, et al. Online static voltage stability monitoring for power systems using PMU data[J]. Mathematical Problems in Engineering, 2020, 2020(1): 6667842.

[145]Momoh JA, El-Hawary ME, Adapa R. A review of selected optimal power flow literature to 1993. II. Newton, linear programming and interior point methods[J]. IEEE Transactions on Power Systems, 1999, 14(1): 105-111.

[146]Zhang W, Li F, Tolbert LM. Review of reactive power planning: Objectives, constraints, and algorithms[J]. IEEE Transactions on Power Systems, 2007, 22(4): 2177-2186.

[147]Zhang W, Tolbert LM. Survey of reactive power planning methods[C]. In: Proceedings of the IEEE Power Engineering Society General Meeting, June 2005: 1430-1440.

[148]Pandya K. A survey of optimal power flow method[J]. Journal of Theoretical and Applied Information Technology, 2008, 4(5): 450-458.

[149]Qiu Z, Deconinck G, Belmans R. A literature survey of optimal power flow problems in the electricity market context[C]. In: Proceedings of the IEEE/PES Power Systems Conference and Exposition, March 2009: 1-6.

[150]Xia X, Elaiw AM. Optimal dynamic economic dispatch of generation: A review[J]. Electric Power Systems Research, 2010, 80(8): 975-986.

[151]Frank S, Steponavice I, Rebennack S. Optimal power flow: A bibliographic survey I[J]. Energy Systems, 2012, 3(3): 221-258.

[152]Bienstock D. Progress on solving power flow problems[J]. Optima, 2013, 93(1-8): 1390.

[153]Frank S, Rebennack S. An introduction to optimal power flow: Theory, formulation, and examples[J]. IIE Transactions, 2016, 48(12): 1172-1197.

[154]Abdi H, Beigvand SD, Scala ML. A review of optimal power flow studies applied to smart grids and microgrids[J]. Renewable and Sustainable Energy Reviews, 2017, 71: 742-766.

[155]Mohagheghi E, Alramlawi M, Gabash A, Li P. A survey of real-time optimal power flow[J]. Energies, 2018, 11(11): 3142.

[156]Street A, Moreira A, Arroyo JM. Energy and reserve scheduling under a joint generation and transmission security criterion: An adjustable robust optimization approach[J]. IEEE Transactions on Power Systems, 2014, 29(1): 3-14.

[157]Dong F, Chowdhury BH, Crow ML, Acar L. Improving voltage stability by reactive power reserve management[J]. IEEE Transactions on Power Systems, 2005, 20(1): 338-345.

[158]Capitanescu F. Assessing reactive power reserves with respect to operating constraints and voltage stability[J]. IEEE Transactions on Power Systems, 2011, 26(4): 2224-2234.

[159]Rabiee A, Soroudi A. Stochastic multiperiod OPF model of power systems with HVDC-connected intermittent wind power generation[J]. IEEE Transactions on Power Delivery, 2014, 29(1): 336-344.

[160]De M, Goswami SK. Optimal reactive power procurement with voltage stability consideration in deregulated power system[J]. IEEE Transactions on Power Systems, 2014, 29(5): 2078-2086.

[161]Rabiee A, Parniani M. Voltage security constrained multi-period optimal reactive power flow using benders and optimality condition decompositions[J]. IEEE Transactions on Power Systems, 2013, 28(2): 696-708.

[162]Ding Y, Xie M, Wu Q, Østergaard J. Development of energy and reserve pre-dispatch and re-dispatch models for real-time price risk and reliability assessment[J]. IET Generation, Transmission & Distribution, 2014, 8(7): 1338-1345.

[163]Martínez-Lacañina PJ, Martínez-Ramos JL, de la Villa-Jaén A, Marano-Marcolini A. DC corrective optimal power flow based on generator and branch outages modelled as fictitious nodal injections[J]. IET Generation, Transmission & Distribution, 2013, 8(3): 401-409.

[164]Tamimi B, Vaez-Zadeh S. An optimal pricing scheme in electricity markets considering voltage security cost[J]. IEEE Transactions on Power Systems, 2008, 23(2): 451-459.

[165]Zhang S, et al. A critical review of data-driven transient stability assessment of power systems: principles, prospects and challenges[J]. Energies, 2021, 14(21): 7238.

[166]Thukaram D, Yesuratnam G. Optimal reactive power dispatch in a large power system with AC–DC and FACTS controllers[J]. IET Generation, Transmission & Distribution, 2008, 2(1): 71-81.

[167]Jabr RA. Adjustable robust OPF with renewable energy sources[J]. IEEE Transactions on Power Systems, 2013, 28(4): 4742-4751.

[168]Shi L, Wang C, Yao L, Ni Y, Bazargan M. Optimal power flow solution incorporating wind power[J]. IEEE Systems Journal, 2012, 6(2): 233-241.

[169]Bhagwan Das D, Patvardhan C. Useful multi-objective hybrid evolutionary approach to optimal power flow[J]. IEE Proceedings - Generation, Transmission and Distribution, 2003, 150(3): 275.

[170]Zabaiou T, Dessaint L, Kamwa I. Preventive control approach for voltage stability improvement using voltage stability constrained optimal power flow based on static line voltage stability indices[J]. IET Generation, Transmission & Distribution, 2014, 8(5): 924-934.

[171]Alhasawi FB, Milanovic JV. Techno-economic contribution of FACTS devices to the operation of power systems with high level of wind power integration[J]. IEEE Transactions on Power Systems, 2012, 27(3): 1414-1421.

[172]Li Y, Zhang S, Li Y, et al. PMU measurements-based short-term voltage stability assessment of power systems via deep transfer learning[J]. IEEE Transactions on Instrumentation and Measurement, 2023, 72: 2526111.

[173]Li P, Wei H, Li B, Yang Y. Eigenvalue-optimisation-based optimal power flow with small-signal stability constraints[J]. IET Generation, Transmission & Distribution, 2013, 7(5): 440-450.

[174]Gayme D, Topcu U. Optimal power flow with large-scale storage integration[J]. IEEE Transactions on Power Systems, 2013, 28(2): 709-717.

[175]Momoh JA, Dias LG, Guo SX, Adapa R. Economic operation and planning of multi-area interconnected power systems[J]. IEEE Transactions on Power Systems, 1995, 10(2): 1044-1053.

[176]Liu C, Wang J, Fu Y, Koritarov V. Multi-area optimal power flow with changeable transmission topology[J]. IET Generation, Transmission & Distribution, 2014, 8(6): 1082-1089.

[177]Zhang R, Dong ZY, Xu Y, Wong KP, Lai M. Hybrid computation of corrective security-constrained optimal power flow problems[J]. IET Generation, Transmission & Distribution, 2014, 8(6): 995-1006.

[178]Li Y, et al. A two-stage multi-objective optimal power flow algorithm for hybrid AC/DC grids with VSC-HVDC[C]//2017 IEEE Power & Energy Society General Meeting. IEEE, 2017: 1-5.

[179]S. Pukhrem, M. Basu, M. F. Conlon, et al. Enhanced Network Voltage Management Techniques Under the Proliferation of Rooftop Solar PV Installation in Low-Voltage Distribution Network[J]. IEEE Journal of Emerging and Selected Topics in Power Electronics, vol. 5, no. 2, pp. 681-694, June 2017.

[180]WANG L, XIE X R, JIANG Q R, et al. Investigation of SSR in practical DFIG-based wind farms connected to a series-compensated power system[J]. IEEE Transactions on Power System, 2015, 30(5):2772-2779.

[181]LIU H K, XIE X R, HE J B, et al. Damping DFIG-associated SSR by adding subsynchronous suppression filters to DFIG converter controllers[C]‖Proceedings of the 2016 IEEE Power and Energy Society General Meeting (PESGM). Boston, USA: IEEE, 2016: 1-5.

[182]GHAFOURI M, KARAAGAC U, KARIMI H, et al. An LQR controller for damping of subsynchronous interaction in DFIG-based wind farms[J]. IEEE Transactions on Power Systems, 2017, 32(6):4934-4942.

[183]LIU H K, XIE X R, LI Y, et al. Mitigation of SSR by embedding subsynchronous notch filters into DFIG converter controllers[J]. IET Generation, Transmission & Distribution, 2017, 11 (11): 2888-2896.

[184]ZHANG X G, XIA D N, FU Z C, et al. An Improved feedforward control method considering PLL dynamics to improve weak grid stability of grid-connected inverters[J]. IEEE Transactions on Industry Applications, 2018, 54(5): 5143-5151.

[185]MA J, SHEN Y Q. DFIG active damping control strategy based on remodeling of multiple energy branches[J]. IEEE Transactions on Power Electronics, 2021, 36(4): 4169-4186.

[186]Cao J, Zhang M, Li Y. A review of data-driven short-term voltage stability assessment of power systems: Concept, principle, and challenges[J]. Mathematical Problems in Engineering, 2021, 2021(1): 5920244.

[187]Park S, Chen W, Mak T W K, et al. Compact optimization learning for AC optimal power flow[J]. IEEE Transactions on Power Systems, 2023.

[188]Gao M, Yu J, Yang Z, et al. A physics-guided graph convolution neural network for optimal power flow[J]. IEEE Transactions on Power Systems, 2023, 39(1): 380-390.